# E-Governance in India: Definitions, Challenges and Solutions

Puneet Kumar
Assistant Professor
FASC, Mody University
Lakshmangarh, Rajasthan

Dharminder Kumar
Professor & Chairman
Dept. of CSE
GJUST, Hisar, Haryana

Narendra Kumar
Assistant Professor
FASC, Mody University
Lakshmangarh, Rajasthan

## ABSTRACT
The Government of India is transcending from traditional *modus operandi* of governance towards technological involvement in the process of governance. Currently, the Government of India is in the transition phase and seamlessly unleashing the power of ICT in governance. The government is spending an enormous amount of finances in deployment of e-governance, but, are these efforts are going in the appropriate direction and leads towards intended results? What do the people percept from the concept of e-governance? What is the global perspective about perception of e-governance? What are the major challenges confronting the deployment of e-governance? In this attempt the authors have made an attempt to riposte aforesaid issues. Moreover, the authors have also suggested some plausible suggestions which may help in successful and sustainable deployment of e-governance in India.

## 1. INTRODUCTION
In 1970 the Government of India (GoI) has established Department of Electronics and subsequently in 1977 GoI has taken first major step towards implementation of e-governance by establishment of National Informatics Centre (NIC). By 1980 most of the government offices were equipped with computers but their role was confined up to word processing. With the span of time and advent of ICT, the GoI has taken a remarkable step for fostering e-governance by launching the national satellite based network NICNET in 1987 followed by District Information System of the National Informatics Centre (DISNIC) and [1] NICNET was the first government informatics network across the world equipped with facilities like TELNET, FTP, internet along with database services (GISTNIC and MEDLARS). Up to 1990, NICNET has extended its extent from state headquarters to district headquarters. In year 2000, the GoI has established Ministry of Information Technology and identified minimum 12-points minimum agenda for e-governance. Finally in year 2006 the GoI has launched National e-Governance Plan (NeGP) with various Mission Mode Projects (MMPs) to automate essential mundane tasks. This paper attempts to define the meaning of e-governance in national and international perspective. Furthermore it tries to demystify about major challenges in implementation of e-governance in India. It also suggests some inferences to triumph the success of e-governance especially in context of India.

## 2. DEFINING E-GOVERNANCE
The U.S. e-government Act, 2002 delineates e-government as *"The use by the Government of web-based Internet applications and other information technologies, combined with processes that implement these technologies, to enhance the access to and delivery of Government information and services to the public, other agencies, and other Government entities or bring about improvements in Government operations that may include effectiveness, efficiency, service quality, or transformation;"*. Whereas the European Union defines it as *"e-Government is the use of Information and Communication Technologies in public administrations combined with organisational change and new skills in order to improve public services and democratic processes"* [2]. Further, [3,4, 5] defined e-governance as a tool for leveraging the potential of Information and Communication Technology (ICT) for improving effectiveness of government activities, strengthening democratic process which led to more empowered citizens and more transparent government offices. In the current era e-government has transformed from being *"just another office tool"* to a powerful utility for innovation, change and a tool for rejuvenating public sector [6]. It is [7] pertinently mentioned that e-governance and e-government are being used as a synonym in Indian perspective. e-Governance can be defined as; use of ICT in government in ways that either alters governance structures or processes in ways that are not feasible without ICT and/or create new governance structures or processes that were heretofore not possible without ICT and/or reify heretofore theoretical ideas or issues in normative governance. Nevertheless, normative governance means ideology of governance which encompasses transparency, accountability, integrity, honesty, impartiality, efficiency etc. in terms of possessing, delivering and enabling. It's a form of e-business in governance which encompasses the processes and structures tangled in delivery of electronic services to the public [8]. Further, [9] defines that, e-government is the modernization of processes and functions of the government by inculcating ICT tools whereas citizens are treated as passive recipients of digital information and services. Nevertheless e-governance is a decisional process which involves ICT in governance with the objective of wider participation and deeper involvement of citizens, institution, NGOs and other companies. It has been also visualized that [10] *"A transparent smart e-Governance with seamless access, secure and authentic flow of information crossing the interdepartmental barrier and providing a fair and unbiased service to the citizen."*

## 3. MAJOR CHALLENGES
The government is expanding an enormous amount on cultivating the culture of e-governance through NeGP but despite of that results are not overwhelming. Although there are islands of success in the area of e-governance but still there are certain areas which are unexplored or inadequately explored. On the basis of the survey; 85% failure rate for e-governance projects has been noticed across the world whereas in year 2009 alone; the worldwide expenditure on the technology abided by the government was 428.38 billion US dollars [11]. The documents [12, 13, 14, 15] had identified major impediments and challenges in successful implementation of e-governance in India. According to the report inadequate planning, leadership failures, deficiency in finances, lack of motivation and awareness, dearth of citizen centric nature of applications, poor cooperation among bureaucrats and people at local level, lack of trust, miserable technical design which endures lack of interoperability among distinct e-governance applications and underutilization of ICT infrastructure resources are the major obstacles in successful implementation of e-governance in India. Although at international level various novel technologies are being used for facilitating e-governance such as semantic web *(Germany)*, virtual reality *(Sweden)*, voice recognition *(Florida)*, Twitter *(USA)*, cloud computing *(UK)* etc. but interoperability in e-governance applications is the prime vision across the world [16]. The authors [17, 18, 19] also advocated the development of community centric e-governance applications after identification of local needs, conditions and demands. The author [20] delineated about parameters for analysing impact of e-governance services and states that quality of service delivered by e-governance and cost abided by the citizen for availing e-services are the major factors responsible for sustainability of any e-governance programme. It has been also advised that a survey must be conducted for analysing degree of relevance in order to mitigate the occurrence of failure. Moreover, it also recommends an impact assessment analysis for every e-governance project and suggests incorporation of impact assessment and evaluation module within the application. Furthermore [21] also accentuated on impact assessment of e-services and suggested a model for the measurement of e-government and [22] also suggested an indicator model for evaluation of transformation effects, infrastructure investments, political and sociological effect, economic and sustainability impacts with reference to e-governance.

The study [23] advocated the use of Service Oriented Architecture (SOA) in e-governance for data management, consolidation of numerous offered services, plummet the cost of IT infrastructure by integrating services, enhancing reusability and facilitating interoperability among various e-governance applications. The effort [24] has identified challenges in designing e-governance applications mentioned as: ability to handle heterogeneity of applications, ability to achieve interoperability to accomplish seamless exchange of data, skill to handle scalability so that new applications can be used as an addendum in the existing systems and can handle growing number of users. The ability to attain high level of data abstraction to maintain privacy of data was also identified as a primary concern. To eradicate these problems the authors have suggested the use of SOA in designing e-governance applications. As interoperability is directly proportional to reusability of the e-governance applications [25] therefore it also requires an apt attention.

The government is expanding an enormous amount on cultivating the culture of e-governance through NeGP but despite of that results are not overwhelming. Although there are islands of success in the area of e-governance but still

## 4. SUGGESTIONS

a. A hybrid approach needs to be adopted for enhancing interoperability among e-governance applications which will encompass [26]centralized approach for document management, knowledge management, file management, grievance management etc. and distributed approach for land registration, building plans, vehicle registration, criminal and crime information etcetera.

b. The Cloud computing is also becoming a big force to enhance delivery of services related to e-governance [27, 28]. The cloud computing is not only a tool for cost reduction but also it helps in; enabling new services, improving education system and creating new jobs/ opportunities [29]. The paper [30, 31] also affirms the usage of cloud for e-governance applications to exaggerate the availability of information, diminish the cost of ICT infrastructure and enhancing interoperability between applications. The government of Japan has established "Kasumigaseki Cloud" to deliver public services to its citizens and according to government of Singapore; it is a major source of economic development [32].

c. The e-governance initiatives in the rural areas should be taken by identifying and analysing the grass root realities [33]. The [34, 35] states that the strategy devised for the implementation of e-governance should be comprehensive; an approach should be citizen centric and should follow multiple channels of communication for dissemination of e-services.

d. The government should also focus on devising appropriate, feasible, distinct and effective capacity building mechanisms for various stakeholders *viz* bureaucrats, rural masses, urban masses, elected representatives etc.

## 5. CONCLUSION

Although, the Government of India is acclaiming its success in the area of e-governance but the scenario at the grass root level is not overwhelming and seamless efforts of the government seems to go in vain. Therefore it's the high time to adopt and imbibe mentioned preventive measures in order to conquer intended objectives of e-governance. Further, the government should learn some lessons from world leaders in the segment of e-governance like Australia. The Australia is continuously focussing in standardization of data, interoperability in e-governance applications and discovering more innovative avenues for delivering e-services effectively.